\documentclass[]{spie}  
\pdfoutput=1

\usepackage[]{graphicx}

\def\arcmin{\hbox{$^\prime$}}

\def\deg{\hbox{$^\circ$}}

\def\lae{\mathrel{\raise .4ex\hbox{\rlap{$<$}\lower 1.2ex\hbox{$\sim$}}}}
\def\gae{\mathrel{\raise .4ex\hbox{\rlap{$>$}\lower 1.2ex\hbox{$\sim$}}}}

\title{Polarimetry with a soft x-ray spectrometer} 

\author{Herman L. Marshall\supit{a}
\skiplinehalf
\supit{a}MIT Kavli Institute, Cambridge, MA, USA
}

\authorinfo{Further author information: E-mail: hermanm@space.mit.edu, Telephone: 1 617 253 8573} 

\begin{document} 
  \maketitle 

\begin{abstract}
An approach for measuring linear X-ray polarization over a broad-band using conventional
spectroscopic optics is described.  
A set of multilayer-coated flats reflect the dispersed X-rays to the instrument detectors.
The intensity variation as a function of energy and position angle is measured to
determine three Stokes parameters: I, Q, and U.
By laterally grading the multilayer optics and matching the dispersion of the gratings,
one may take advantage of high multilayer reflectivities and achieve modulation factors over 80\%
over the entire 0.2 to 0.8 keV band.
A sample design is shown that could be used with a small orbiting mission.  
\end{abstract}



\section{INTRODUCTION}
\label{sec:intro}  

Although polarization is a fundamental property of electromagnetic radiation,
there have been no astronomical measurements of polarization in the X-ray
band in over 30 yr.  There has been no lack of workable concepts, including
simple designs such as the Polarimeter for Low Energy X-ray Astrophysical
Sources (PLEXAS) that was proposed by Marshall et al.\ (2003) to use
multilayer-coated mirrors tuned to 0.25 keV as Bragg reflectors.\cite{plexas}
An excellent review of the history and prospects for astronomical
polarimetry in the 0.1-10 keV band is presented by Weisskopf et al. (2006)\cite{weisskopf06}.
Weisskopf et al.\ rightly argue that the PLEXAS
design had a narrow bandpass, reducing its attractiveness
for astrophysical observations because one expects polarization to be energy
dependent, so a wide bandpass is desired.

Marshall (2007\cite{2007SPIE.6688E..31M}; hereafter, Paper I)
described a method to overcome this limitation by using transmission gratings
to disperse in the incoming X-rays.  Then, a multilayer-coated reflector would
be used to modulate the signal in a way depending on the polarization.
Two approaches were suggested.  The first was a simple modification of
a dispersive spectrometer, such as a grating spectrometer proposed for
Constellation-X (Con-X).  The second approach was intended to provide
a means for obtaining polarimetric data by inserting optics forward of the
focal plane of a long focal length telescope such as XEUS.

Similar to one of the approaches suggested in
Paper I, this paper is limited to the
case where the polarization is determined
by measuring spectral intensity along different dispersed spectra.
In this case, we consider a configuration that would be
appropriate for a small orbiting observatory.

\section{A Soft X-ray Polarizing Spectrometer}

\subsection{The Spectrometer}

As in Paper I,
the approach to this polarimeter design was inspired by a new blazed transmission grating
design, called the Critical Angle Transmission (CAT) grating \cite{Heilmann:08}, and the
corresponding application to the Con-X mission in the design of
a transmission grating spectrometer\cite{flanagan07}.
For details of the Con-X mission, see the Con-X web site at
{\tt http://constellation.gsfc.nasa.gov/}.  This type
of grating can provide very high efficiency in first order in the soft
X-ray band.  For a spectrometer, one places detectors on the Rowland
torus, which is slightly ahead of the telescope's imaging surface.

As in the design for Con-X, using a sector of the primary allows one
to take advantage of the asymmetric scattering profile of the mirror
surfaces.  As noted by Cash (1987, and references therein)\cite{},
the scattering profile is much broader in the plane that contains the
optical axis (radially away from the mirror surface) than in the azimuthal
direction.  Thus, the images produced using a small sector of the
mirror will be narrow in one dimension and broad in the other.  By
arranging the gratings so that the dispersion direction is along the
direction of tight imaging, one obtains better spectral resolution
than if the entire aperture was used.  We exploit this feature to provide
modest spectral resolution in each of many sectors, even with mirrors
with 0.5-1.5\arcmin\ half-power diameter imaging.

There are several factors involved in setting the telescope
focal length: the grating period, $P$; the mirror
imaging size (in one dimension), $\delta \theta$; and the shortest wavelength of
interest, $\lambda_{min}$.  For a given focal length, $F$, and placing
the gratings at, say, $0.9 F$ from the polarizer, we find that the
spectral resolution is given by $\frac{0.9 \lambda_{min}}{P \delta \theta}$. 
For this study, we target a spectral
resolution of $\sim$ 200 at the shortest wavelengths of interest, near
1.5 nm (0.8 keV), so that we may obtain
a spectral resolution that is $> \times 2$ better than that
provided by the multilayer coatings.
This criterion gives $\delta \theta = 14$\deg\ in one dimension.
The mirror half-power diameter can be 4-10$\times$ larger
than this value, due to the larger in-plane mirror scattering.
Thus, mirrors with average imaging of order 1-2\arcmin\ would be
acceptable for this project.

The system focal length is set only by the requirement that the
detector pixels are at least $\times 4$ smaller than the spectral
resolution element, $F \delta \theta$.  For $25 \mu$ pixel
sizes, then $F > 1.5$ m, so the telescope can be very short.
Effective area is a premium in polarimetry, due to the expected
low polarization signals.  For a graze angle of 1\deg, needed to reflect
0.8 keV X-rays off Al mirrors at $> 90$\% efficiency,
the mirror diameter could be 28 cm for $F = $2 m.
Assuming an inner diameter of 10 cm and an average reflectivity
to two reflections of about 70\%, the effective area of the mirror
would be about 200 cm$^2$.

\subsection{The Polarizer}

\begin{figure}
   \begin{center}
   \begin{tabular}{c}
   \includegraphics*[height=13cm,angle=0]{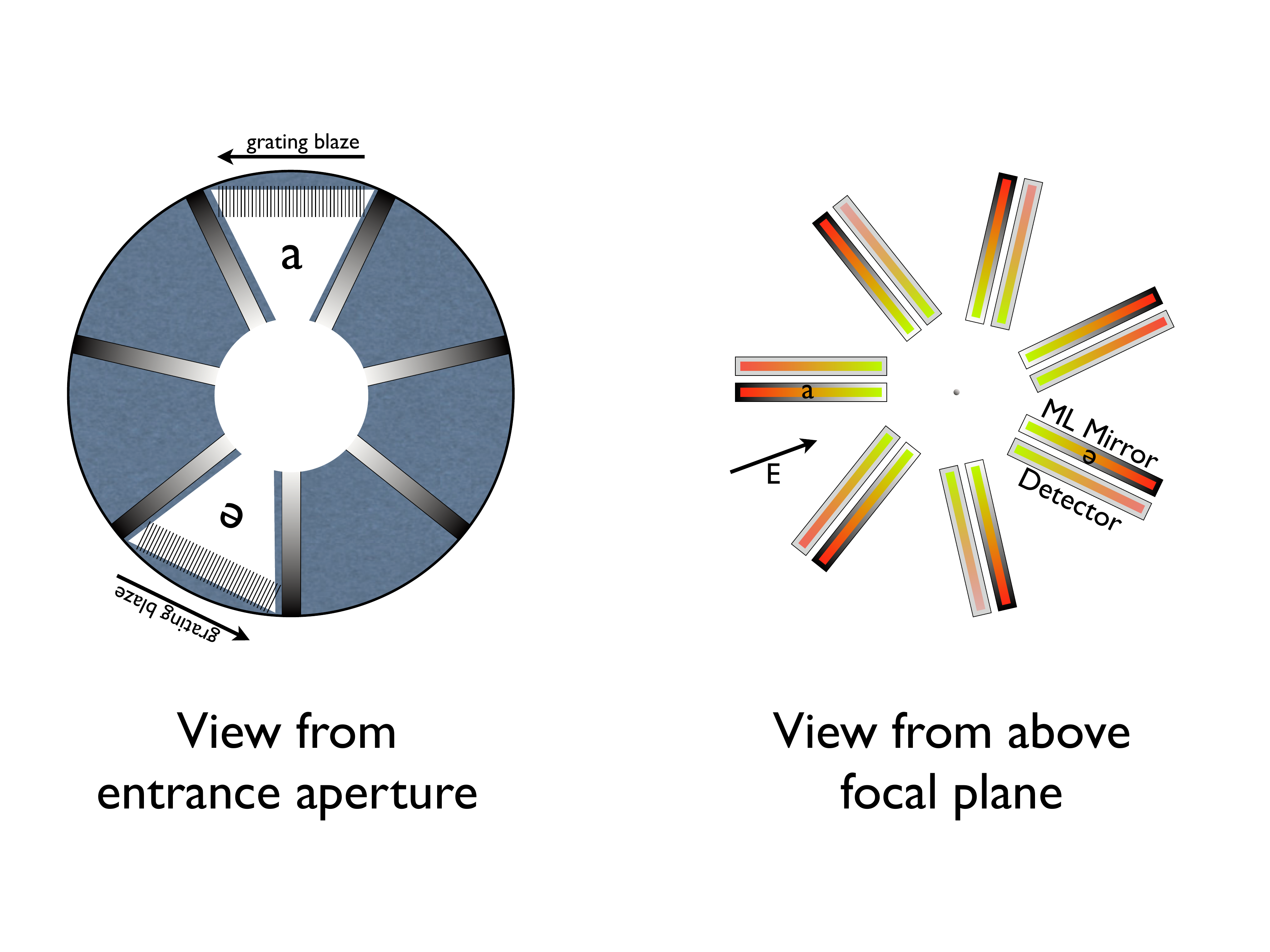}
   \end{tabular}
   \end{center}
 \caption{
{\it Left:} View from the entrance aperture of a mirror with seven sectors.  Gratings are placed
behind the mirror with the grating bars oriented along the average radius to the mirror axis.
The gratings then disperse X-rays in the directions given for sectors {\bf a} and {\bf e}, as
shown (lettering proceeding clockwise from the top central sector).  The gratings are blazed,
so the directions to the orders with highest efficiency are shown for two sectors.
 {\it Right:} The view above the detectors and polarizers.  Spectra from the gratings are
 incident on multilayer-coated flats that are tilted along the dispersion axis which contains
 the zeroth order (dot in center).  The angle of the tilt is the same for all mirrors and always
 redirects the X-rays to the adjacent detector in a clockwise direction.  The mirror
 assembly would appear something like a pinwheel.
The detectors are tilted in the opposite direction -- always oriented counterclockwise, to face
the corresponding mirror.  For a mirror graze angle of 40\deg, the detector normals would be
80\deg\ from the optical axis.  Mirrors {\bf a} and {\bf b} receive X-rays from the
corresponding sector shown in the left side of the figure and all zeroth order light converges
to the spot at the center for an on-axis source.
For polarized X-rays with an electric vector as shown, the intensity at a given wavelength can
vary from detector to detector, as shown schematically here.  In this case, long wavlengths
at the perimeter are highly modulated with position angle while short wavelengths, closer
to the optical axis, are not modulated at all, simulating a wavelength-dependent polarization
fraction.
 }
\label{fig:layout08}
\end{figure}

A CAT-based spectrometer becomes a polarimeter
by placing multilayer-coated flats on the Rowland circle instead
redirecting and polarizing the spectra.  The flats are tilted about
the spectral dispersion axis by an angle $\theta$.
For this study, the graze angle, $\theta$ is 40\deg.
The detectors are then
oriented toward the mirrors at an angle $90-2*\theta = 10$\deg\ to the
plane perpendicular to the optical axis of the telescope.
Fig.~\ref{fig:layout08} shows how the optics might look.
In this case, the entrance aperture is divided into seven sectors,
with gratings aligned to each other in each sector and along the
average radial direction to the optical axis.  This approach
generates seven spectra that are reflected by seven polarizing
flats to seven detectors.

The multilayer spacing, $d$, on the polarizing flats would
vary laterally (along the dispersion) in order to provide optimal reflection
at graze angle $\theta$.
Such optimization is obtained by equating the wavelength from
the linearized grating equation, $m \lambda = P x/D_R$, to that of a multilayer,
$\lambda = 2 d \sin \theta$, giving

\begin{equation}
\label{eq:dspace}
d = \frac{Px}{2 m D_R \sin \theta} \,  ,
\end{equation}
where $x$ is the distance along the dispersion
from 0th order, $D_R$ is the Rowland diameter
(the distance between the gratings and the usual detector plane),
and $m$ is the grating order.
Thus, the multilayer period varies linearly with $x$, providing high reflectivity
in a narrow bandpass at large graze angles.  For a given order,
the spectrometer's gratings direct a narrow bandpass to any given position
on the flat, so the bandpass of the multilayer can be similarly narrow.
The bandpass of the multilayer decreases approximately as $1/N$ for a large
number of layers, $N$, so if the spectrometer has a resolution better than 200,
say, then $N \gae 100$.  Large $N$ also improves the multilayer reflectivity.

At Brewster's angle, $\theta = 45$\deg, reflectivity is minimized when the $E$-vector
is in the plane containing the incident ray and the surface normal ($p$-polarized)
and maximized when the $E$-vector is in the surface plane ($s$-polarized).
The polarization position angle (PA) is the average orientation in sky coordinates
of the $E$-vector for the incoming X-rays.
Sampling at least 3 PAs is required in order to measure
three Stokes parameters (I, Q, U) uniquely, so one would require at
least three separate detector
systems with accompanying multilayer-coated flats.
For this study and for the sake of redundancy, seven detector-flat combinations
are assumed.

Following Marshall et al.\cite{pkspolar} and Paper I,
the intensity $I_\lambda$ at wavelength $\lambda$ varies with
azimuthal angle $\phi$ as

\begin{equation}
I_\lambda \left({\phi }\right)=\ {I}_\lambda^{0}\left\{{{\frac{1}{2\pi }}\ +\ {P_\lambda^0} M
	\cos \left({2\left[{\phi \rm -\varphi_\lambda }\right]}\right)}\right\}\ \ ,
\end{equation}

\noindent
where $I_\lambda^0$ is directly related to the I Stokes parameter of the source
as a function of wavelength $\lambda$,
$P_\lambda^0$ is the polarization fraction at this
wavelength, $\varphi_\lambda$ is the phase determined by the orientation
of average polarization $E$-vector on the sky (i.e., the polarization PA),
and $M_\lambda$ is the average system modulation factor.
The modulation factor is a function of wavelength, given by
$M = \frac{R_s-R_p}{R_s+R_p}$,
where $R_s$ and $R_p$ are the
reflectivities for the $s$ and $p$ polarizations, respectively.
The Stokes Q and U parameters can be derived from $P_0$ and
$\varphi$ as a function of wavelength as long as there are three or
more position angles that do not differ from each other by 180\deg.

\subsection{System Throughput and Sensitivity}

Using the Center for X-ray Optics web page
({\tt http://henke.lbl.gov/optical\_constants/multi2.html}),
multilayer efficiencies were computed for a small range of materials for
many of the relevant graze angles and corresponding $d$ values.
For this work, $N$ was set to 200 and the interdiffusion
thickness was set to 0.3 nm, as has been achieved by several
groups\cite{windt04,1998ApOpt..37.1873M,1996ApOpt..35.5134M}.
Table~\ref{tab:mlspecs} gives a sampling of the results, where
$r$ is the peak reflectivity to unpolarized X-rays.  The thickness
of the first material is $0.4 d$, while that of the second material
is $0.6 d$ for each calculation.  All modulation
factors are larger than 85\%.

\begin{table}[h]
\caption{Sample Multilayer Specifications} 
\label{tab:mlspecs}
\begin{center}       
\begin{tabular}{|c|lllll|} 
\hline
$\lambda$	&	$E$	&	$d$	&	$r$	&	$M$	&	Materials	\\
(nm)	&	(keV)	&	(nm)	&		&		&		\\
\hline											
1.61	&	0.77	&	1.25	&	0.013	&	0.939	&	W/Si	\\
1.80	&	0.69	&	1.40	&	0.025	&	0.939	&	Ni/Mg	\\
1.93	&	0.64	&	1.50	&	0.039	&	0.941	&	Ni/Mg	\\
2.06	&	0.60	&	1.60	&	0.081	&	0.925	&	Ni/Mg	\\
2.57	&	0.48	&	2.00	&	0.094	&	0.915	&	Ni/Mg	\\
3.09	&	0.40	&	2.40	&	0.106	&	0.913	&	Ni/Mg	\\
3.15	&	0.39	&	2.45	&	0.258	&	0.856	&	Cr/Sc	\\
3.99	&	0.31	&	3.10	&	0.128	&	0.990	&	Cr/Sc	\\
4.37	&	0.284	&	3.40	&	0.123	&	0.903	&	Cr/Sc	\\
4.63	&	0.268	&	3.60	&	0.232	&	0.884	&	C/Cr	\\
7.71	&	0.161	&	6.00	&	0.123	&	0.902	&	C/Cr	\\\hline 
\end{tabular}
\end{center}
\end{table} 

\begin{figure}
   \begin{center}
   \begin{tabular}{c}
   \includegraphics*[height=15cm,angle=0]{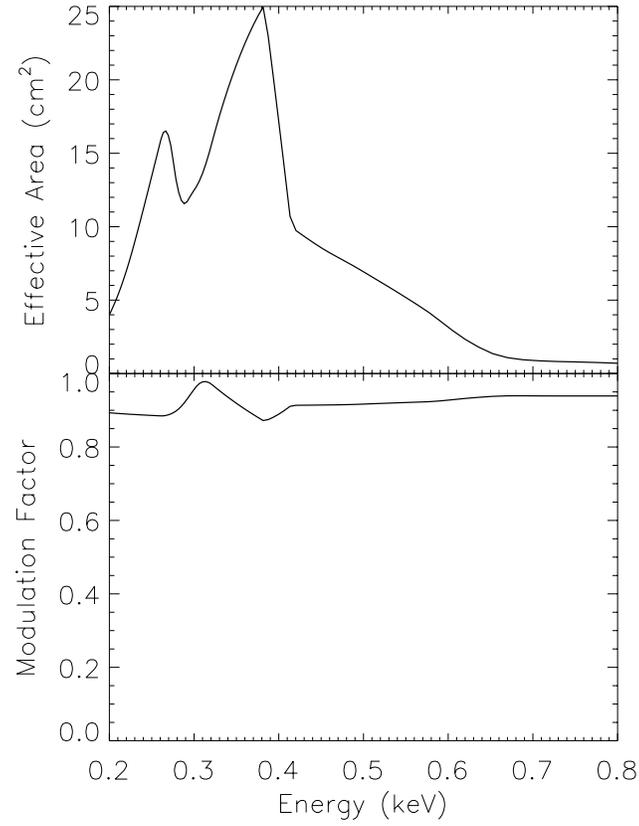}
   \end{tabular}
   \end{center}
 \caption{
{\it Top:} 
{\it Bottom:} Polarization modulation factor as a function of energy across the
bandpass of the instrument.}
\label{fig:effarea}
\end{figure}

For the purposes of this analysis, a CCD detector is assumed to have a thin
Al overlayer to act as an optical blocking filter.
The resulting effective area is shown in
Fig.~\ref{fig:effarea}, along with the modulation factor, $M$.

\begin{figure}
   \begin{center}
   \begin{tabular}{c}
   \includegraphics*[height=15cm,angle=90]{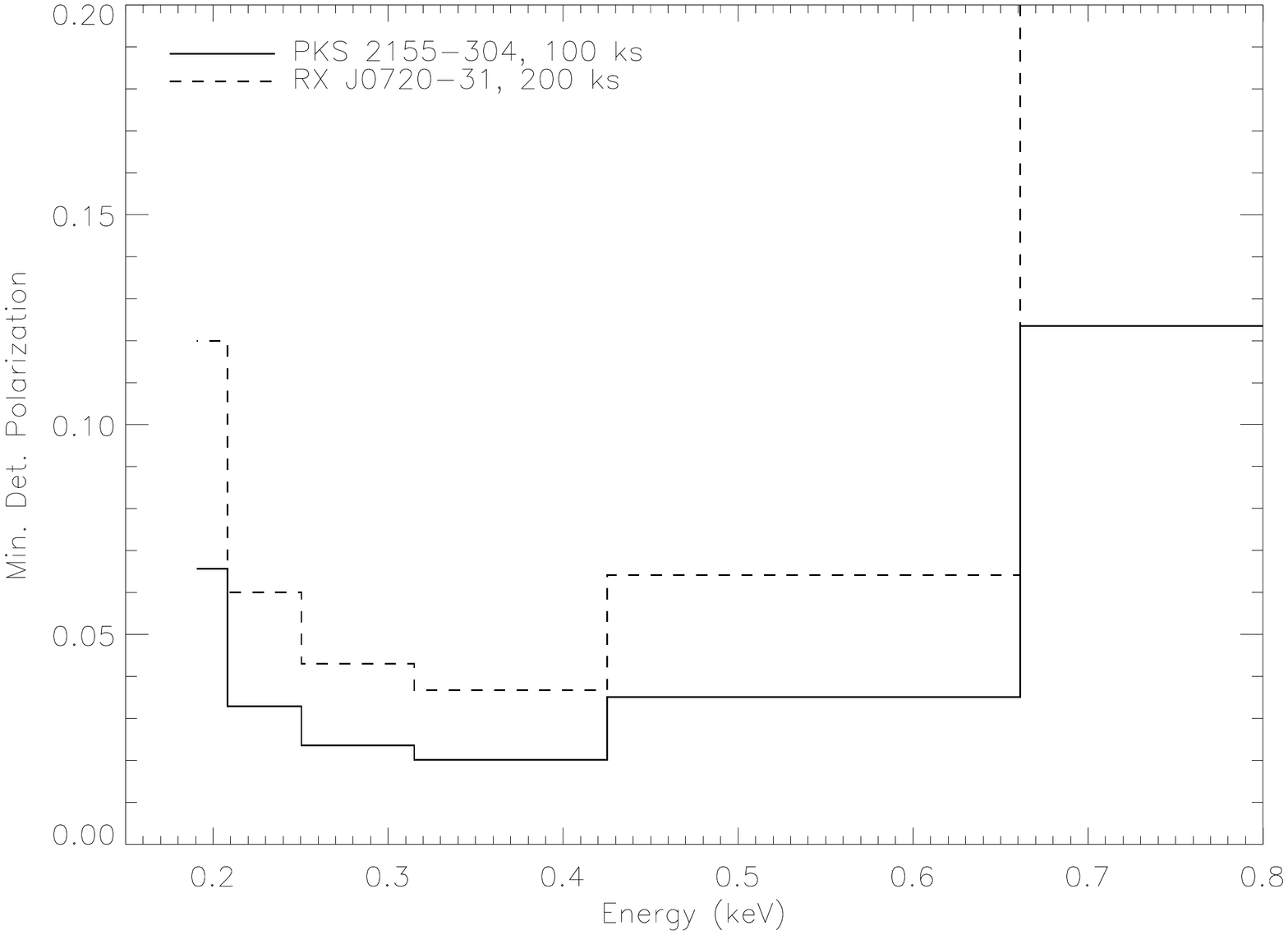}
   \end{tabular}
   \end{center}
 \caption{Minimum detectable polarization as a function of energy across the
bandpass of the instrument for two different possible observations.
The solid line shows how we could detect linear polarization at a level
of 15-20\% across the
entire energy band from 0.2 to 0.8 keV for PKS 2155-304 in 100 ks.
For RX J0720-31, spectroscopy allows one to obtain the polarization
below, in, and above absorption features.
}
\label{fig:mdp}
\end{figure}

The effective area estimate can be used to predict the minimum
detectable polarization\cite{weisskopf06} (MDP)
for potential targets.  As described by Marshall et al\cite{plexas},
extragalactic sources such as the BL Lac object
PKS~2155$-$304 are expected to be highly polarized in the X-ray band.
In a 100,000 s observation of PKS~2155$-$304,
this instrument could detect polarizations of 3-7\% in
each of four bandpasses 1 nm wide from 2 to 6 nm (0.21 to 0.62 keV).
Fig.~\ref{fig:mdp} shows the results graphically.
In addition, isolated neutron stars such as RX J0720.4$-$3125 are expected to polarized
due to effects of photon propagation in strong magnetic fields\cite{weisskopf06}.  In
this case, it is worthwhile to obtain phase-resolved polarization data.  In a 200,000 s
observation with this instrument, one could reach MDPs of 5-8\% over the 2-5 nm
band (in 1 nm wide bands) for each of 10 phase bins.

There are many combinations of elements and multilayer coating parameters that were
not investigated for this preliminary study, so one might expect to improve upon the
mulilayer-coating reflectivities.
It may be possible
to superpolish the substrates to obtain 0.1 nm roughness\cite{2001SPIE.4506..113K},
which would increase reflectivities significantly.

\section{A Test and Calibration Facility}

\begin{figure}
   \begin{center}
   \begin{tabular}{c}
   \includegraphics*[height=10cm,angle=0, viewport= 177 170 850 600]{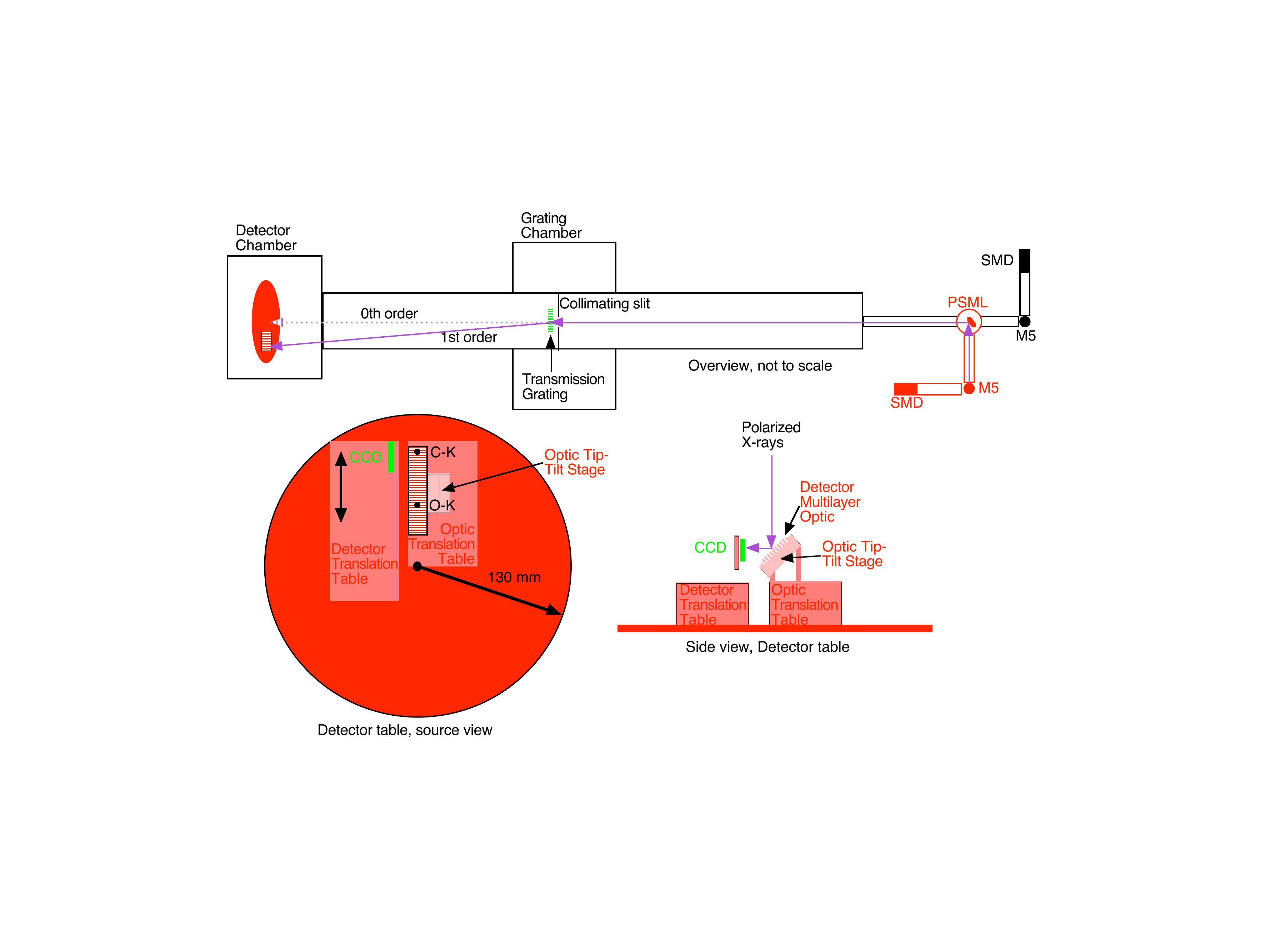}
   \end{tabular}
   \end{center}
 \caption{Schematic of the MIT 17 m beamline after proposed
modifications for soft X-ray polarimetry prototyping. Items in red would be
added to the system.  The SMDs are source monitor detectors -- proportional
counters that view the associated Manson Model 5 source (M5).  The
X-ray beam is shown in light purple; a new source would be used with the
polarized source multilayer (PSML) angled at 45$\deg$ to pass polarized X-rays.
The bottom two views are
subject to detailed mechanical design and stage dimensions.
The dispersed,
first order X-rays go to the detector multilayer optic, angled at 45$\deg$ to
the incoming X-rays.  The CCD faces the multilayer optic on a translation
table that moves the CCD to the location appropriate to the dispersion
by the grating (Table~\ref{tab-lines}).
The CCD stage and its associated multilayer optic are mounted
on a platform and can be rotated $\pm 180\deg$.
}
\label{fig:polnlab}
\end{figure}

Drs.\ R.\ Heilmann and N.\ Schulz
are currently working to fully recommission the
X-ray grating evaluation facility (X-GEF), a 17 m beamline that was developed
for testing transmission gratings fabricated at MIT for
the {\it Chandra} project\cite{1994SPIE.2280..257D}.
In addition to plans to test CAT and other types of gratings, we
have proposed to add equipment to the facility so that it
may be used for testing and calibrating components that
could be used in a soft X-ray polarimeter.  Fig.~\ref{fig:polnlab}
shows how it might appear when operational.

\begin{table}
\begin{center}
\caption{Polarization Source Emission Lines}
\begin{tabular}[t]{|c|c|c|} \hline
\label{tab-lines}
Emission Line & Energy & Multilayer Spacing \\
& (keV) & (nm) \\
\hline
C-K & 0.277 & 3.165\\
Ti-L & 0.452 & 1.940\\
O-K & 0.525 & 1.670\\
Cr-L & 0.573 & 1.530\\
Fe-L & 0.705 & 1.244 \\
\hline
\end{tabular}
\end{center}
\end{table}

Equipment to be added to X-GEF would include a new X-ray source
with multilayer coated optics to produce highly polarized X-rays at each of
five X-ray emission lines (see Table~\ref{tab-lines}), a
grating rotation stage to simulate telescope
rotation, and a multilayer mirror and CCD combination acting as the
polarimeter.  The detector and its associated mirror would be mounted
on a rotation stage to match the rotation of the grating so that the
polarization modulation factor can be measured.  The detector's
multilayer-coated mirror is challenging to fabricate but two vendors
have provided quotes.

\acknowledgments     
 
Support for this work was provided by the National Aeronautics and
Space Administration through the Smithsonian Astrophysical Observatory
contract SV3-73016 to MIT for Support of the Chandra X-Ray Center,
which is operated by the Smithsonian Astrophysical Observatory for and
on behalf of the National Aeronautics Space Administration under contract
NAS8-03060.


\bibliography{polarimeter}   
\bibliographystyle{spiebib}   

\end{document}